# A Conjecture about a "vision" model for blind men.

# L. A. Amarante Ribeiro

**Retired Associate Professor of Physics Department, ICEx, UFMG**

lincoln@coltec.ufmg.br <mailto:Lincoln@coltec.ufmg.br>

## Abstract

In this work we present a "vision" model for blind people, preferentially for those ones of birth origen. The model should be implemented using the avaiable computational resources at the moment. The model uses a simulation of an optical image by an appropriate sonorous "image" according to certain rules.

It is an old dream of mankind to solve the problem of the human blindness in a general way and of the congenital blindness in particular, to get a method that allows to the faulty visual to have an insight of the images that surround him and to have a common language with those which see in the traditional way. The rise of computers and modern sensors gives us a great hope that we can build a sophisticated equipment to overcome that limitation of the human being, the visual deficiency. At least an attempt in that direction is already on progress[1]. In this work we will present a possible solution for the problem. Although the model is good for all the well-known types of blindness, it better adapts to the congenital blindness, perhaps due to the long time that will be worn-out in the learning of the method by a blind man. That is at least what we preliminarily suppose. The method that we will denominate Vision-sonic is based on the following fundamental point: It is well known by the psychologists, synesists and even for lay that there is a perceptive sharpness of another active senses in the blind men by the lack of the sight of the vision. In general they are much more sensitive to several aspects of the audition as height, timbre and intensity than a person which has perfect ownership of its visual abilities. By this reason we had the idea of trying to use this aspect to supply the vision deficiency. We intend to use sounds to simulate the faulty of an image in the brain. This can be obtained, for example in the following way: a digital camera gets a digitalized image of the landscape which the person wants to observe. Soon after a file of images is generated, Bitmap say (BMP) of the photo. The digital file, this way generated, is swept in a computer through a special software that has the function of transforming each pixel of the image in a sound information recorded in a file WAV , say according to a rule that we expose here. There are several ways of doing the correspondence image-sound with advantages and disadvantages at least now they won't be discussed here. In a first method we make a correspondence which can be : for the red (R)the note la (A), the note fa (F) for green t(G) the note re (D) for blue (B). So each visual pixel RGB would be equivalent to a sound chord AFD. The choice of very temperate musical notes is arbitrary. We could choose any frequency for the sound that represents each color following a convenience pointed by psychologists. Chosen the notes or frequencies, the heights of the sound correspond to the basic colors and the intensities of the notes of a chord, the representation of the color tone, they would correspond to the several proportions of RGB in the tone of the pixel. A second method consists of doing the same correspondence between the basic colors and the musical notes. However the proportions with each color enters in a certain color tone would be simulated by a duration of the sound and not by its intensity. So we would have for each tonality, the emission of three notes in succession, instead of a chord. A third naïve method, but perhaps the one of easier learning would be to use simply the seven musical notes to simulate the seven colors of the rainbow and we leave the resolution sideways in different tones. Only the experience can decide for one of these methods. Once implemented the technological aspects of the transformation of image files in the corresponding sound files, we would passe to the other phase: the one of a child's instruction in the recognition of the representative sounds of the several images. We could begin for example, showing him a yellow card with a blue circle of a certain diameter. We would describe the content of the image for the child and we make him to hear the corresponding sound. Soon after we would modify the color of the circle for red and we repeated the procedure with other colors, until he memorize the most important colors of the spectrum. Soon after we would change the size of the circle progressively with the same color, so that it is acquired the size notion, we move for another illustrations and so forth being the child intructed to associate a certain group of sounds with the several aspects of an image, such as, its forms, position, dimensions, color, etc. With this we believe that we can build an engine able to do people with visual deficiency come " to see " in a very close future . If the deficiency is not coming from lesions in the responsible cerebral area for the vision, it is possible to excite this area by means of a equipment built with this purpose transforming the sound in electric pulses that they will work the nerves transmitters of the light, substituting the paper of rhodopsin[2]

Conclusive results regarding to a blind man real capacity to interpret the incoming sound signs of an image is beyond the purposes of this article and the only way of test thating that point of view it is through experiences.